\documentclass[twocolumn,showpacs,preprintnumbers,amsmath,amssymb,superscriptaddress]{revtex4}%

\usepackage{graphicx}
\usepackage{dcolumn}
\usepackage{bm}
\usepackage{color}
\usepackage[version=3]{mhchem}

\begin{document}

\preprint{preprint(\today)}
\title{Nodeless Superconductivity and its Evolution with Pressure in the Layered Dirac Semimetal 2M-WS$_{2}$}


\author{Z.~Guguchia}
\email{zurab.guguchia@psi.ch}\affiliation{Laboratory for Muon Spin Spectroscopy, Paul Scherrer Institute, CH-5232 Villigen PSI, Switzerland}

\author{D.~Gawryluk}
\email{dariusz.gawryluk@psi.ch}\thanks{On leave from Institute of Physics, Polish Academy of Sciences, Aleja Lotnikow 32/46, PL-02-668 Warsaw,
Poland.}
\affiliation{Laboratory for Multiscale Materials Experiments, Paul Scherrer Institut, 5232 Villigen PSI, Switzerland}

\author{M.~Brzezinska}
\affiliation{Physik-Institut der Universit\"{a}t Z\"{u}rich, Winterthurerstrasse 190, CH-8057 Z\"{u}rich, Switzerland}
\affiliation{Department of Theoretical Physics, Wroclaw University of Science and Technology, Wyb. Wyspianskiego 27, 50-370 Wroclaw, Poland}

\author{S.S.~Tsirkin}
\affiliation{Physik-Institut der Universit\"{a}t Z\"{u}rich, Winterthurerstrasse 190, CH-8057 Z\"{u}rich, Switzerland}

\author{R.~Khasanov}
\affiliation{Laboratory for Muon Spin Spectroscopy, Paul Scherrer Institute, CH-5232 Villigen PSI, Switzerland}

\author{E.~Pomjakushina}
\affiliation{Laboratory for Multiscale Materials Experiments, Paul Scherrer Institut, 5232 Villigen PSI, Switzerland}

\author{F.O.~von~Rohr}
\affiliation{Department of Chemistry, University of Z\"{u}rich, CH-8057 Z\"{u}rich, Switzerland}
\affiliation{Physik-Institut der Universit\"{a}t Z\"{u}rich, Winterthurerstrasse 190, CH-8057 Z\"{u}rich, Switzerland}

\author{J.~Verezhak}
\affiliation{Laboratory for Muon Spin Spectroscopy, Paul Scherrer Institute, CH-5232 Villigen PSI, Switzerland}

\author{M.Z.~Hasan}
\affiliation{Laboratory for Topological Quantum Matter and Spectroscopy, Department of Physics, Princeton University, Princeton, New Jersey 08544, USA}

\author{T.~Neupert}
\affiliation{Physik-Institut der Universit\"{a}t Z\"{u}rich, Winterthurerstrasse 190, CH-8057 Z\"{u}rich, Switzerland}

\author{H.~Luetkens}
\affiliation{Laboratory for Muon Spin Spectroscopy, Paul Scherrer Institute, CH-5232 Villigen PSI, Switzerland}

\author{A.~Amato}
\affiliation{Laboratory for Muon Spin Spectroscopy, Paul Scherrer Institute, CH-5232 Villigen PSI, Switzerland}

\begin{abstract}
The interaction between superconductivity and band topology can lead to various unconventional  superconducting (SC) states, and represents a new frontier in condensed matter physics research. 
Recently, the transition metal dichalcogenide (TMD) system 2M-WS$_{2}$ has been identified as a Dirac semimetal exhibiting both superconductivity with the highest $T_{\rm c}$ ${\sim}$ 8.5 K among all the TMD materials and  topological surface states with a single Dirac cone. Here we report on muon spin rotation (${\mu}$SR) and density functional theory studies of microscopic SC properties and the electronic structure in 2M-WS$_{2}$ at ambient and under hydrostatic pressures ($p_{\rm max}$ = 1.9 GPa). The SC order parameter in 2M-WS$_{2}$ is determined to have single-gap $s$-wave symmetry. We further show a strong negative pressure effect on $T_{c}$ and on the SC gap ${\Delta}$. This may be partly caused by the pressure induced reduction of the size of the electron pocket around the ${\Gamma}$-point, at which a band inversion appears up to the highest applied pressure. We also find that the superfluid density $n_{s}$ is very weakly affected by pressure. The absence of a strong pressure effect on the superfluid density and the absence of a correlation between $n_{s}$ and $T_{\rm c}$ in 2M-WS$_{2}$, in contrast to the other SC TMDs $T_{\rm d}$-MoTe$_{2}$ and 2H-NbSe$_{2}$, is explained in terms of its location in the optimal (ambient pressure) and above the optimal (under pressure) superconducting regions of the phase diagram and its large distance to the other possible competing or cooperating orders. These results hint towards a complex nature of the superconductivity in TMDs, despite the observed  $s$-wave order parameter. 

\end{abstract}
\maketitle

\section{INTRODUCTION}

The transition metal dichalcogenides (TMDs) are representative layered materials presently attracting very strong interest from research communities working on nanoscience and 2-d layered systems  \cite{Soluyanov,Kaminski,Xu,Ali1,QiCava,Qian,Zhang,Ugeda,Bozin,SunY,Hasan1,Le,KivelsonCDW,Yokoya}. The TMDs share the \ce{MX2} chemical formula, with M being a transition metal (M = Ti, Zr, Hf, V, Nb, Ta, Mo, W or Re) and X being a chalcogen (X = S, Se, or Te). TMDs are obtained in various crystal structures 1T, 2H, and 1T$^{\prime}$ with different stackings of the individual \ce{MX2} layers. A number of ground states have been observed in these systems including charge-density waves, superconductivity, semiconductors and topological metals. 
New physics findings in TMDs include strong excitonic states, novel phenomena from valley splitting, Dirac cone dispersion and topological effects, and more \cite{Soluyanov,Ali1}.\\

Layered materials, with highly anisotropic electronic properties have been found to be potential hosts for unconventional superconductivity. TMDs exhibit many features strikingly similar to those reported in other unconventional superconductors. Superconductivity can be observed under pressure as for the pristine 1T polytype TMD materials which are not superconducting at ambient pressure, but exhibit charge density wave (CDW) states, instead  \cite{Sipos}. The metallic 2H polytypes have two, or more, Fermi surfaces and saddle bands, allowing for dual orderings, which can be coexisting CDW and SC orderings, two superconducting gaps, two CDW gaps, and possibly even pseudogaps above the onset temperature of CDW orderings, as reported in some cuprate superconductors \cite{Tidman,Thompson,Naito}. Upon application of pressure, the CDW ground state is suppressed and the superconducting transition temperature is enhanced. A dome-like superconducting phase was induced by electrostatic gating in the semiconducting 2H-MoSe$_{2}$, 2H-MoS$_{2}$, 2H-MoTe$_{2}$ and 2H-WS$_{2}$ with a $T_{\rm c}$ of up to 10 K \cite{Iwasa2012,Iwasa2015,Klemm}. It was also shown that using field effect gating, one can induce superconductivity
in monolayer semiconducting TMD WS$_{2}$ \cite{JLu}, and one could access an unprecedented doping range revealing a rich set of competing electronic phases ranging from band insulator, via superconductor, to  a reentrant insulator at high doping. Superconductivity with a very low $T_{\rm c}$ ${\sim}$ 0.1 K was also reported in the Weyl semimetal MoTe$_{2}$ which exhibits a low temperature orthorhombic $T_{\rm d}$ structure \cite{QiCava}. However, under the application of pressure, the orthorhombic structure transforms into the  1T$^{\prime}$ structure and $T_{\rm c}$  is enhanced to about 8 K.  Recently, we found a correlation between the superfluid density at $T$ = 0 and the transition temperature $T_{\rm c}$ in the type-II Weyl semimetal $T_{d}$-MoTe$_{2}$ \cite{GuguchiaMoTe2} and the deviation from the linear correlation in 2H-NbSe$_{2}$ under pressure \cite{GuguchiaNbSe2}. These observations resemble the ones found earlier in cuprates \cite{Uemura1,Uemura3,Uemura7,Uemura4,EmeryKivelson,Shengelaya}  and Fe-based superconductors \cite{Luetkens,Khasanov2008,Carlo,GuguchiaPRB}.
These examples demonstrate that the TMDs exhibit rather unconventional superconducting and normal state properties.\\

\begin{figure*}[t!]
\centering
\includegraphics[width=1.0\linewidth]{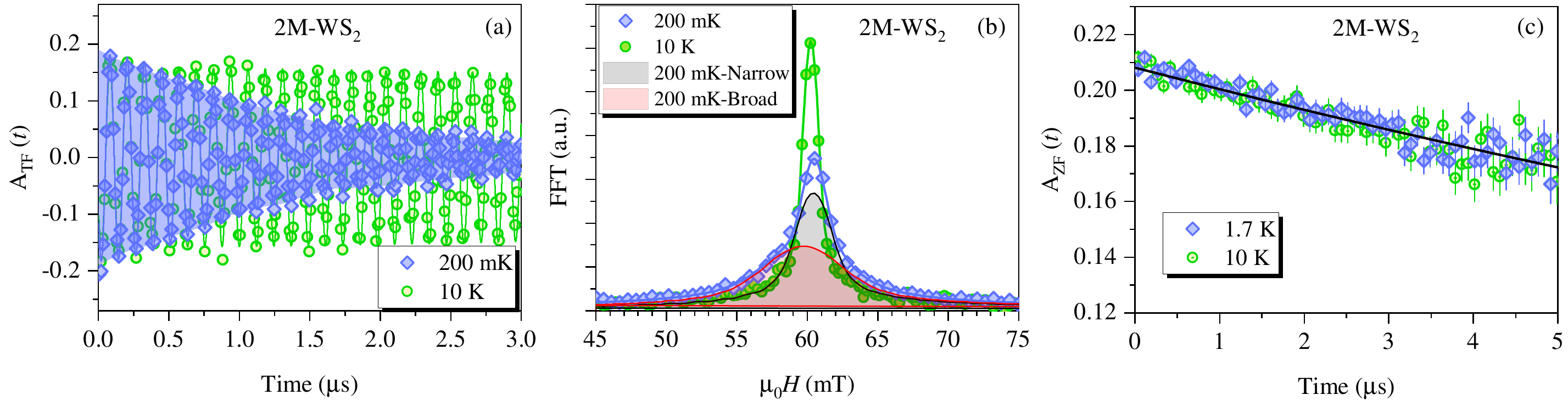}
\vspace{-0.7cm}
\caption{\textbf{Transverse-field (TF) and Zero-field (ZF) ${\mu}$SR time spectra for 2M-WS$_{2}$.} 
(a) The TF spectra, above and below $T_{\rm c}$ are shown. The solid lines represent fits to the data by means of Eq.~1; 
(b) The Fourier transforms of the ${\mu}$SR time spectra shown in panel (a).   
(c) ZF ${\mu}$SR time spectra for 2M-WS$_{2}$ recorded above and below $T_{\rm c}$. The line represents the fit to the data of the combination of Lorentzian and Gaussian Kubo-Toyabe depolarization function reflecting the field distribution at the muon site created by nuclear dipole moments and randomly oriented diluted weak local electronic moments.}
\label{fig1}
\end{figure*}

 Recently, superconductivity with $T_{\rm c}$ ${\sim}$ 8.8 K was reported in the new compound 2M-WS$_{2}$, which is constructed from 1T$^{\prime}$-WS$_{2}$ monolayers \cite{YFang}. 2M-WS$_{2}$ has the highest $T_{\rm c}$ among all TMD materials at ambient pressure. The system possesses a semimetallic band structure. The crystal structure of 2M-WS$_{2}$ has a different stacking of 1T$^{\prime}$-WS$_{2}$ monolayers along the $a$-axis, compared to $T_{\rm d}$-MoTe$_{2}$ and 1T$^{\prime}$-WTe$_{2}$. In addition to superconductivity, topological surface states with a single Dirac cone were theoretically predicted for 2M-WS$_{2}$. Hence, this system represents one of the few examples of a material with both superconductivity and a topologically non-trivial surface states. The possible occurrence of topological superconductivity (TSC) in 2M-WS$_{2}$ after fine tuning of some parameters is also discussed \cite{YFang}. TSCs are materials with unique electronic states consisting of a full pairing gap in the bulk and gapless surface states composed of Majorana fermions (MFs) \cite{Hosur,Ando,Grushin}. Up to now, the only known properties of the superconducting state in 2M-WS$_{2}$ are the critical temperatures and fields \cite{YFang}. Thus, a thorough exploration of superconductivity in 2M-WS$_{2}$ from both experimental and theoretical perspectives is imperative.

   To further study superconductivity and its nature in 2M-WS$_{2}$, it is critical to measure 
the superconducting order parameter at the microscopic level. Thus, we conducted ambient and high-pressure \cite{GuguchiaPressure, Andreica, MaisuradzePC, GuguchiaNature} muon spin relaxation/rotation (${\mu}$SR)  measurements of the magnetic penetration depth $\lambda$ in 2M-WS$_{2}$.  This quantity is one of the fundamental parameters of a superconductor, since it is related to the superconducting carrier density $n_{s}$ via 1/${\lambda}^{2}$ = $\mu_{0}$$e^{2}$$n_{s}/m^{*}$ [where $m^{*}$ is the effective mass; see also Eq. (2)].  The ${\mu}$SR technique provides a powerful tool to measure ${\lambda}$ in the vortex state of type II superconductors in the bulk of the sample \cite{Sonier}, in contrast to many techniques that probe  ${\lambda}$ only near the surface.


   Here we show that the superconducting order parameter in 2M-WS$_{2}$ possesses single-gap $s$-wave symmetry, which is consistent with the presence of dominant hole-type charge carriers, as deduced from Hall effect measurements.  Moreover, we observed a strong negative pressure effect on the critical temperature $T_{c}$ and the superconducting gap ${\Delta}$ as well as on the size of the electron pocket around the ${\Gamma}$-point, at which a band inversion appears up to the highest applied pressure. However, the pressure effect on the superfluid density in 2M-WS$_{2}$ is very weak. We find that the ratio $T_{\rm c}$/$n_{s}$  is located in the Uemura plot close to those of other superconducting TMD systems, as $T_{\rm d}$-MoTe$_{2}$ and 2H-NbSe$_{2}$, and those of other unconventional superconductors. This suggests that 2M-WS$_{2}$ exhibits rather unconventional superconducting properties. The absence of a strong pressure effect on the superfluid density and its correlation  with $T_{\rm c}$, which was previously found in  $T_{\rm d}$-MoTe$_{2}$ and 2H-NbSe$_{2}$, is explained in terms of the absence of competing or cooperating order to superconductivity, as  2M-WS$_{2}$ exhibits optimal superconducting conditions already at ambient pressure. By application of pressure we are pushing the system away from any possible competing/cooperating order. Our findings, therefore, pose a challenge for understanding the underlying physics in these layered TMDs and might ultimately lead to a better understanding of generic aspects of non-BCS behaviors in unconventional superconductors.

\section{Results}
\subsection{Magnetic Penetration Depth and the Superconducting Gap Symmetry}

 Figure 1a exhibits transverse-field (TF) muon-time spectra for 2M-WS$_{2}$ measured in an applied magnetic field of $\mu_0H =$ 60 mT above ($T$ = 10 K) and below ($T$ = 200 mK) the superconducting (SC) transition temperature $T_{{\rm c}}$. Above $T_{{\rm c}}$ the oscillations reflecting the muon-spin precession show a small damping due to the random local fields from the nuclear magnetic moments. Below $T_{{\rm c}}$ the relaxation rate strongly increases upon decreasing the temperature due to the presence of a nonuniform local magnetic field distribution resulting from the formation of the flux-line lattice (FLL) in the Shubnikov phase. By carefully examining the TF-${\mu}$SR data, we find that the sample consists of a superconducting and a non-superconducting volume fraction. To visualise these two fractions of the sample, we report on Fig. 1b the Fourier transforms of the ${\mu}$SR time spectra, shown in Fig. 1a. At $T$ = 200 mK the narrow signal around ${\mu}_{0}$$H_{\rm ext}$ = 60 mT, originates from the non-superconducting part of the sample, while the broad signal with a first moment ${\mu}_{0}$$H_{int}$ ${\textless}$ ${\mu}_{0}$$H_{\rm ext}$, marked by the solid arrow in Fig. 1b, arises from the superconducting part of the sample. The estimated superconducting volume fraction from the ${\mu}$SR data is ${\sim}$ 45 ${\%}$. It is interesting to investigate whether magnetism is present either in the superconducting or non-superconducting fractions of the sample. In order to search for magnetism, we have carried out zero-field (ZF) ${\mu}$SR experiments above and below $T_{{\rm c}}$, which are shown in Fig. 1c. No sign of static magnetism could be detected in the ZF time spectra down to 1.7 K. Moreover, the ZF relaxation rate is small and does not change between 10 K and 1.7 K. ZF-${\mu}$SR spectra are well described by the combination of a Lorentzian and a Gaussian Kubo-Toyabe depolarization function \cite{Toyabe,GuguchiaPRB2016}, reflecting the field distribution at the muon site created by the nuclear dipole moments and randomly oriented diluted weak local electronic moments. Returning to the discussion of the TF-${\mu}$SR data, we observe a strong diamagnetic shift of the internal magnetic field ${\mu}_{0}$$H_{\rm int}$ sensed by the muons below $T_{{\rm c}}$. This is evident in Fig. 2a, where we plot the temperature dependence of ${\Delta}$$B_{\rm dia}$ = ${\mu}_{0}$($H_{\rm int,SC}$ - $H_{\rm int,NS}$), i.e., the difference between the internal field ${\mu}_{0}$$H_{\rm int,SC}$ measured in the SC fraction and one  ${\mu}_{0}$$H_{\rm int,NS}$ measured in the normal state at $T$ = 10 K. The strong diamagnetic shift along with the negative field dependence of the superconducting depolarization rate ${\sigma}_{{\rm sc}}$ (see Fig. 2b) excludes the occurrence of field induced magnetism in 2M-WS$_{2}$. Note that in some Fe-based and cuprate high temperature superconductors, where field induced magnetism was detected, a paramagnetic shift and a linear increase of  ${\sigma}_{{\rm sc}}$ with magnetic field were observed \cite{KhasanovPRL103,Williams2010,SonierPRL2011}. The absence of magnetism in zero-field or under applied magnetic fields implies that the increase of the TF relaxation rate below $T_{{\rm c}}$ is solely arising from the FLL in the superconducting fraction. The strong superconducting response from nearly 50 ${\%}$ of the sample points to the bulk superconductivity in 2M-WS$_{2}$. We note that the diamagnetic moment, which we deduce from the bulk magnetization measurements suggest a 100 ${\%}$ SC volume in the sample, which is consistent with the previous report \cite{YFang}. However, bulk magnetisation measures the diamagnetic screening of the whole sample where a 100 ${\%}$ screening does not necessarily imply a 100 ${\%}$ SC volume. ${\mu}$SR, being a local probe, directly measures the SC volume fraction within the sample. Thus, in this work we provide a more precise estimate for the SC volume in 2M-WS$_{2}$. The ${\mu}$SR evidence for a phase separation between SC and paramagnetic regions in 2M-WS$_{2}$ is an interesting finding, which might be an intrinsic feature of this system.
 
\begin{figure*}[t!]
\centering
\includegraphics[width=1.0\linewidth]{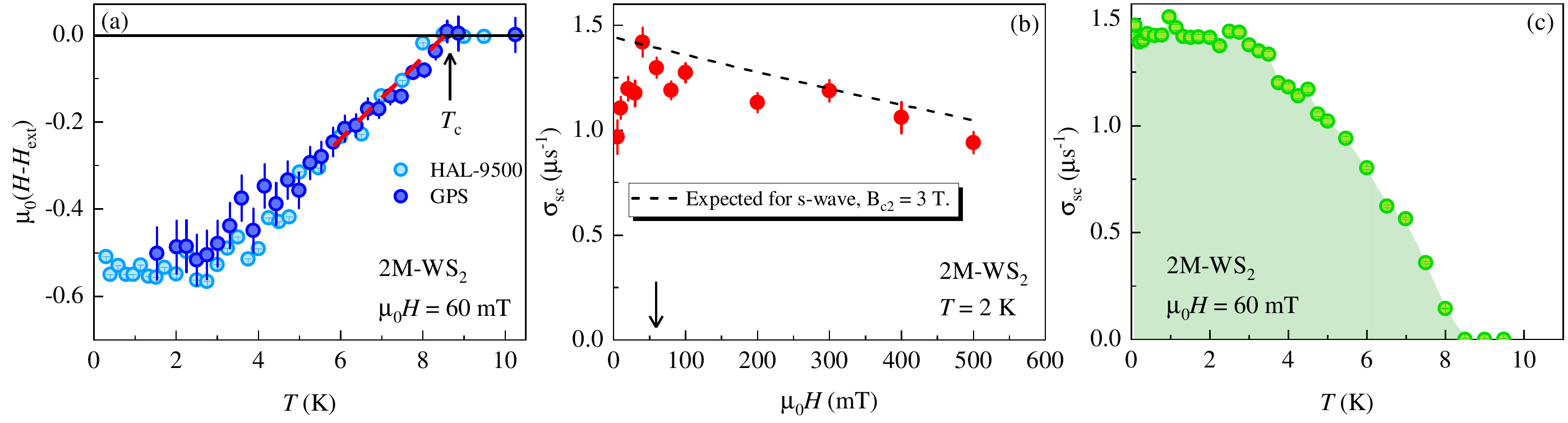}
\vspace{-0.7cm}
\caption{\textbf{Superconducting muon spin depolarization rate ${\sigma}_{\rm sc}$ and the diamagnetic shift for 2M-WS$_{2}$.} 
(a) The temperature dependence of the diamagnetic shift ${\Delta}$$B_{dia}$ = ${\mu}_{0}$($H_{int,SC}$ - $H_{int,NS}$) in an applied magnetic field of ${\mu}_{\rm 0}H = 60$~mT. The arrow denotes $T_{\rm c}$. (b) Field dependence of the  the muon spin relaxation rate ${\sigma}_{\rm sc}$($T$) at 1.6 K. Arrow indicates the field value at which the temperature dependent measurements were carried out. Dashed line shows the theoretical field dependence expected for the $s$-wave superconductor, taking into account the critical field of 3 T. (c) The temperature dependence of ${\sigma}_{\rm sc}$($T$).} 
\label{fig2}
\end{figure*}

From the TF muon time spectra, we determined the Gaussin superconducting relaxation rate ${\sigma}_{{\rm sc}}$ (after subtracting the nuclear contribution), 
which is proportional to the second moment of the field distribution (see Method section). 
Figure 2b shows the obtained field dependence of ${\sigma}_{{\rm sc}}$ at 1.7 K. Each point was obtained by field cooling the sample from above $T_{c}$ to 1.7 K.
As expected from the London model one observes that first ${\sigma}_{{\rm sc}}$ strongly increases with increasing magnetic field until reaching a maximum at 40 mT \cite{Brandt}. Above 40 mT, ${\sigma}_{{\rm sc}}$ decreases with increasing magnetic field. This appears consistent with a behaviour expected for $s$-wave superconductor for an ideal triangular vortex lattice \cite{Brandt}. Taking into account the critical field of 3 T, the theoretical $s$-wave behaviour
has been calculated according to the Brandt formula \cite{Brandt}. It is shown by the dashed line in Fig. 2b, which agrees well with the experimental data.
The second moment of the resulting inhomogeneous field distribution is related to the magnetic penetration depth $\lambda$ by $\left<\Delta B^2\right>\propto \sigma^2_{sc} \propto \lambda^{-4}$ \cite{Brandt}. In order to investigate the symmetry of the superconducting gap, we have therefore derived the temperature-dependent London magnetic penetration depth ${\lambda}(T)$, which is related to the relaxation rate by: 
\begin{equation}
\frac{\sigma_{\rm sc}(T)}{\gamma_{\mu}}=0.06091\frac{\Phi_{0}}{\lambda^{2}(T)}.
\end{equation}
Here, ${\gamma_{\mu}}$ is the gyromagnetic ratio of the muon, and ${\Phi}_{{\rm 0}}$ is the magnetic-flux quantum. 
In order to accurately determine the temperature dependence of $\lambda$, it is important to do the temperature dependent measurements of  ${\sigma}_{{\rm sc}}$ slightly above the maximum of ${\sigma}_{{\rm sc}}(H)$, where Eq. (1) is valid. Thus, the measurements were done with an applied magnetic field of 60 mT. In Fig. 2c, we show the temperature dependence of the muon spin depolarization rate ${\sigma}_{{\rm sc}}$. Below $T_{{\rm c}}$ the relaxation rate ${\sigma}_{{\rm sc}}$ starts to increase from zero with decreasing temperature due to the formation of the FLL and reaches ${\sigma}_{{\rm sc}}$ = 1.4 ${\mu}s$$^{-1}$  at the base-$T$. The temperature dependence of ${\sigma}_{{\rm sc}}$, which reflects the topology of the SC gap, exhibits a saturation at low temperatures. This is consistent with a nodeless superconductor, for which the superfluid density $n_{s}$ ${\propto}$ ${\sigma}_{sc}$  reaches its zero-temperature value exponentially. For a quantitative analysis, we in the local (London) approximation (${\lambda}$ ${\gg}$ ${\xi}$, where ${\xi}$ is the coherence length), calculated ${\lambda}(T)$. The result of the gap analysis is shown in Fig. 3. The data are perfectly compatible with a single $s$-wave gap with a value of ${\Delta}$ ${\simeq}$  1.31(1) meV. 

The London magnetic penetration depth ${\lambda}$ is given as a function of $n_{\rm s}$ \cite{Tinkham}, the effective mass $m^{*}$, ${\xi}$ and the mean free path $l$, according to \\
\begin{equation}
\begin{aligned}
\frac{1}{\lambda^2} = \frac{4\pi n_se^2}{m^*c^2} \cdot  \frac{1}{1 + \xi/l}. 
\end{aligned}
\end{equation} 
For systems close to the clean limit, ${\xi}$/$l$ ${\rightarrow}$ 0, the second term essentially becomes unity, and the simple relation 1/${\lambda}^{2}$ $\propto$ $n_{\rm s}/m^{*}$ holds. Considering the upper critical fields $H_{c2}$ of 2M-WS$_{2}$, as reported in detail by Fang \textit{et al.} \cite{YFang}, we can estimate the in-plane coherence length to be ${\xi}_{ab}$ ${\simeq}$ 8.8 nm at ambient pressure. At ambient pressure, the in-plane mean free path $l$ was estimated to be $l_{ab}$ ${\simeq}$ 200 nm. No estimates are currently available for $l$ under pressure. However, the in-plane $l$ is most probably independent of pressure considering the fact that the pressure mostly reduces the interlayer distance due to the unique anisotropy resulting from the stacking of layers with van der Waals type interactions between them. Thus, in view of the short coherence length and relatively large $l$, we can reliably assume that 2M-WS$_{2}$ lies close to the clean limit \cite{GuguchiaMoTe2,Frandsen}. With this assumption, we obtain the ground-state value $n_{s}/(m^{*}/m_{e}$) ${\simeq}$ 2.8 ${\times}$ 10$^{26}$ m$^{-3}$, where $m_{e}$ is the free electron mass. 

\begin{figure}[b!]
\centering
\includegraphics[width=1.0\linewidth]{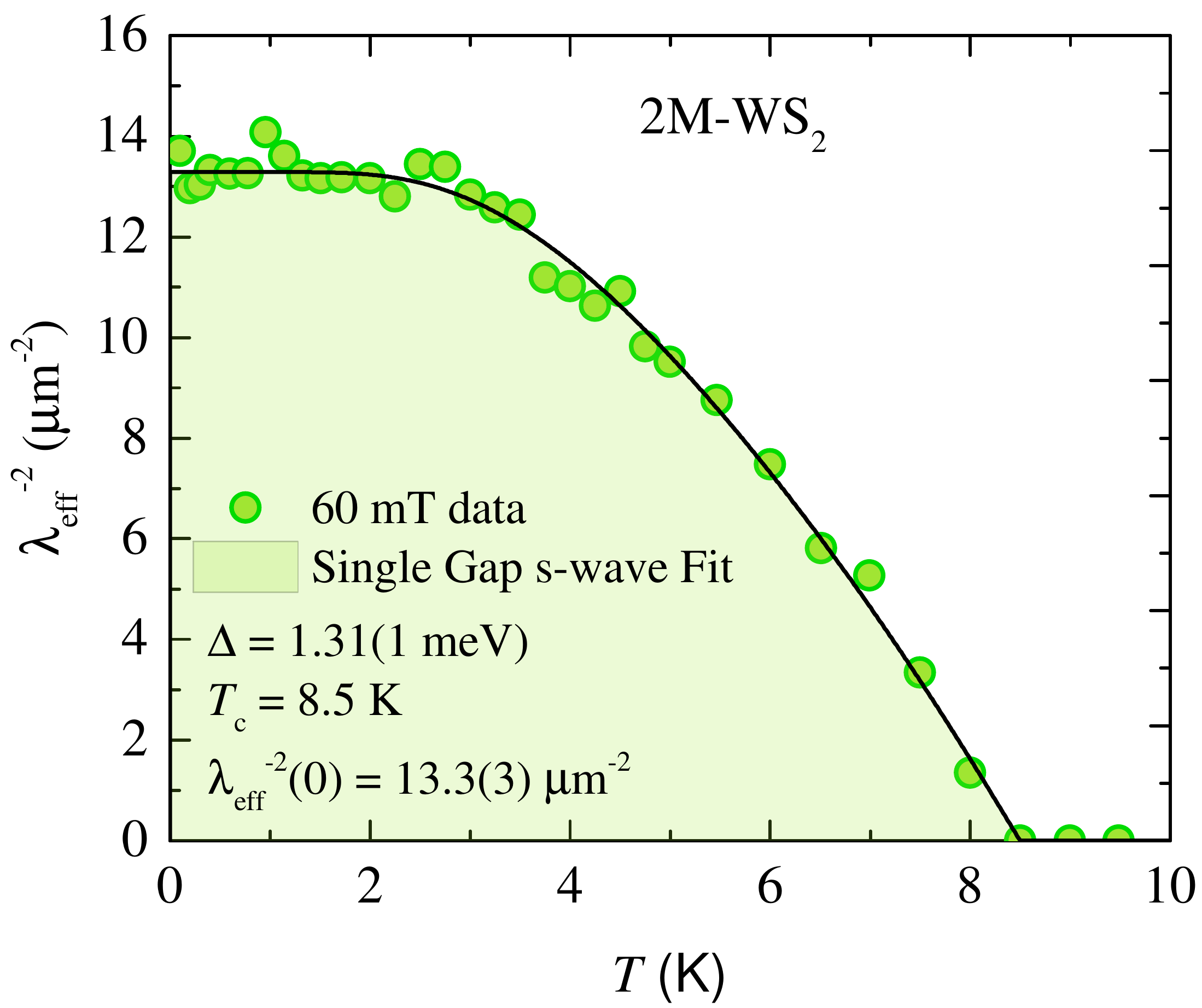}
\vspace{-0.5cm}
\caption{\textbf{Temperature evolution of ${\lambda}^{-2}$ for 2M-WS$_{2}$.}
(a) The temperature dependence of ${\lambda}^{-2}$ measured at ambient pressure. The solid line represents a fit using a single gap $s$-wave model.}
\label{fig3}
\end{figure}

\begin{figure*}[t!]
\centering
\includegraphics[width=1.0\linewidth]{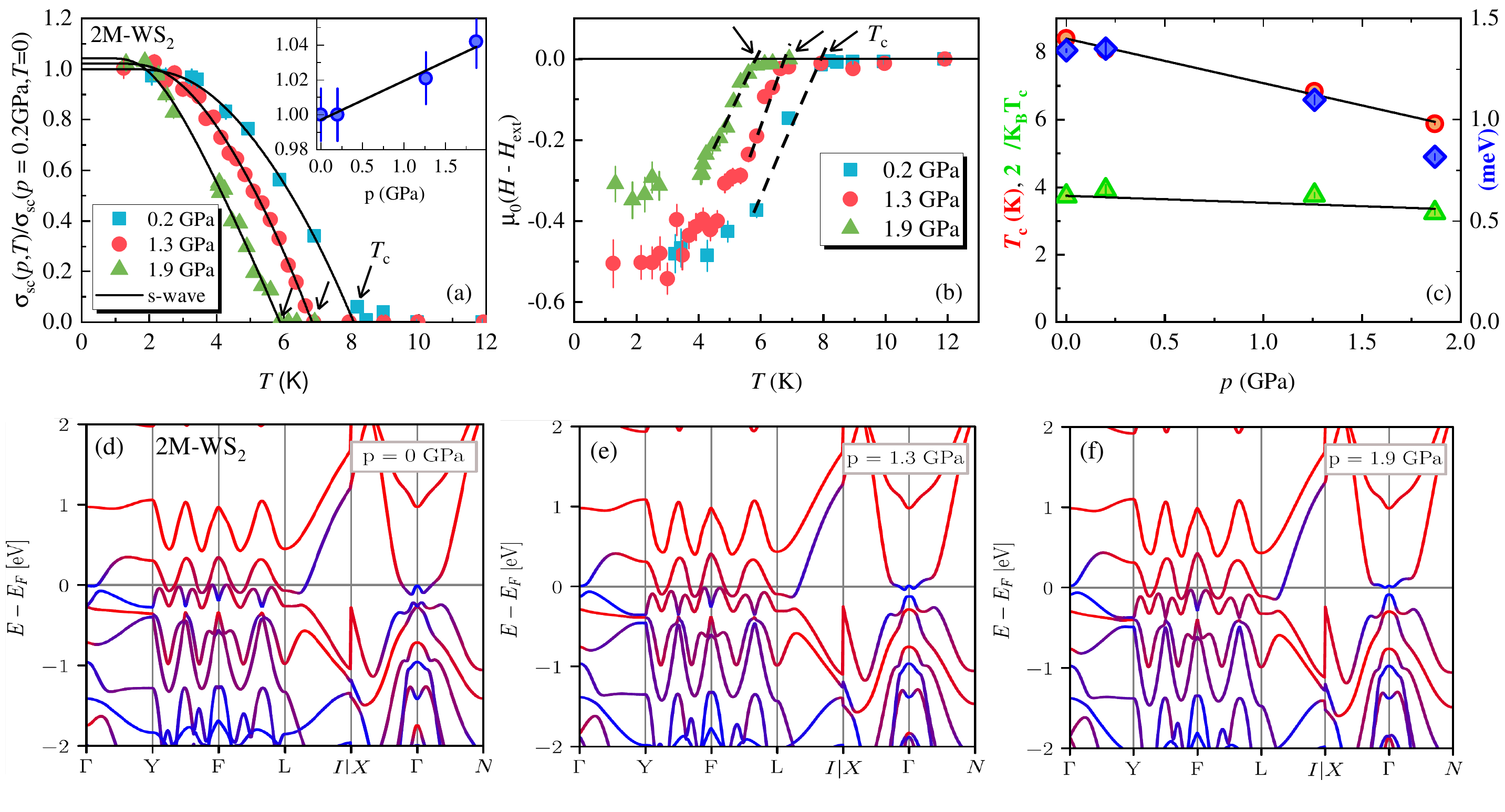}
\vspace{-0.5cm}
\caption{ \textbf{Pressure evolution of various quantities.} 
(a)  The temperature dependence of ${\sigma}_{\rm sc}$($T$), normalized to its value for the lowest applied pressure, under various applied pressures. The inset illustrates the weak positive pressure dependence of ${\sigma}_{\rm sc}$. (b) The temperature dependence of the diamagnetic shift ${\Delta}$$B_{dia}$ = ${\mu}_{0}$($H_{int,SC}$ - $H_{int,NS}$), recorded for various hydrostatic pressures. 
(c) Pressure dependence of the superconducting transition temperature $T_{\rm c}$, the zero-temperature value of the superconducting gap ${\Delta}$ and the ratio 2${\Delta}$/$K_{B}$$T_{\rm c}$ for 2M-WS$_{2}$.  Band structure in the presence of spin-orbit coupling for $p$ = 0 GPa (d), 1.3 GPa (e) and 1.9 GPa (f). Red color corresponds
to $d$ orbitals of W atoms, while the blue color indicates the $p$ orbitals of S atoms.}
\label{fig4}
\end{figure*}

\subsection{Probing superconductivity and band structure as a function of pressure}

To gain further insight into the superconducting properties of 2M-WS$_{2}$, the temperature dependence of the superfluid response was measured under applied pressures up to 1.9 GPa. Moreover, the pressure effect of the electronic structure was also calculated. In Fig. 4a and b, we show the temperature evolution of ${\sigma}_{{\rm sc}}$, normalised to the value at ambient pressure, and the diamagnetic shift ${\Delta}$$B_{\rm dia}$ measured under various hydrostatic pressures. A single gap $s$-wave model describes the ${\sigma}_{{\rm sc}}\left(T\right)$ data very well up to the highest pressure. Remarkably, both $T_{\rm c}$ and ${\Delta}$ decreases substantially upon application of pressure. $T_{\rm c}$ decreases from 8.5 K at $p$ = 0 GPa to  6 K at 1.9 GPa. ${\Delta}$ decreases from 1.31(1) meV at 0 GPa to 0.82(2) meV at 1.9 GPa. However, the analysis of the temperature dependence of ${\sigma}_{{\rm sc}}\left(T\right)$ reflects only a 5 ${\%}$ increase of the zero-temperature value of $n_{s}$. This means that the absolute value of the superfluid density in 2M-WS$_{2}$ is very weakly affected by pressure. The band structures, calculated in the presence of the spin-orbit coupling for ambient pressure and under the applied pressures of $p$ = 1.3 GPa and 1.9 GPa are shown in Figs. 4 (d-f). The band inversion between the $p$-orbital of the S atoms and the $d$-orbital of the W atoms at the ${\Gamma}$-point is found up to the highest investigated pressure. This indicates that the topology of the bands in 2M-WS$_{2}$  remains nontrivial  under pressure. We also find that the electron pockets at the ${\Gamma}$-point shrink upon application of pressure, which might have some interesting consequences on the physical properties of the system. 

\begin{figure*}[t!]
\centering
\includegraphics[width=1.0\linewidth]{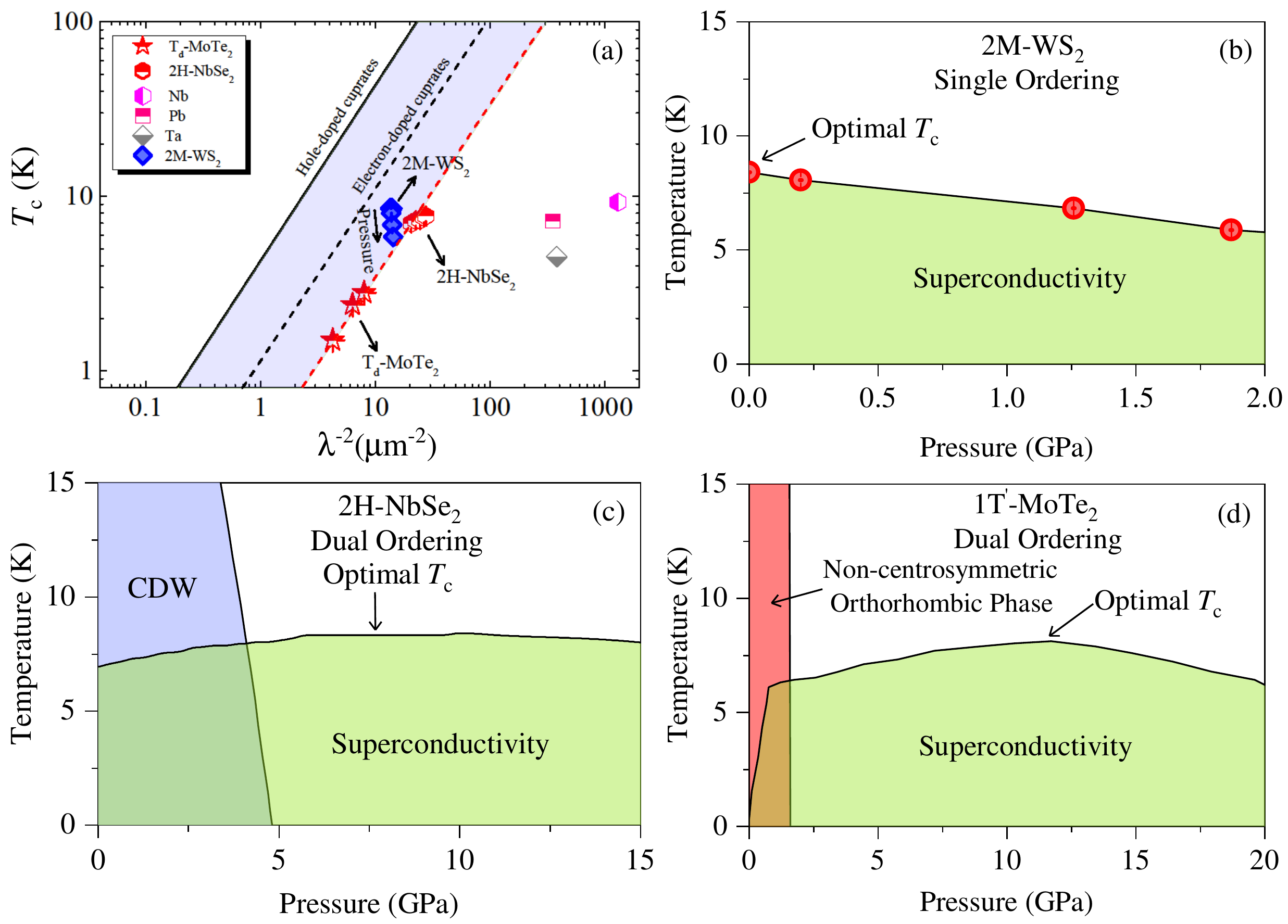}
\vspace{-0.7cm}
\caption{ \textbf{Superfluid Density versus $T_{\rm c}$ and comparison between the phase diagrams.} 
(a) A plot of $T_{\rm c}$ against the ${\lambda}^{-2}(0)$ obtained from our ${\mu}$SR experiments in 
2M-WS$_{2}$, 2H-NbSe$_{2}$ and MoTe$_{2}$. Data points for MoTe$_{2}$ and 2H-NbSe$_{2}$ are taken from Refs. \cite{GuguchiaMoTe2,GuguchiaNbSe2}  The dashed red line represents the linear fit to the MoTe$_{2}$ and 2H-NbSe$_{2}$ data. Uemura relation for hole and electron-doped cuprates are shown as solid \cite{Uemura1,Uemura3,Uemura4} and dashed lines \cite{Shengelaya}, respectively. The points for various conventional BCS superconductors are also shown. The temperature-pressure phase diagram for 2M-WS$_{2}$ (b), showing only the superconducting order. The schematic temperature-pressure phase diagrams for 2H-NbSe$_{2}$ (c)  (after ref. \cite{YFeng}) and MoTe$_{2}$ (after ref. \cite{QiCava}) (d), exhibiting dual ordering.}
\label{fig4}
\end{figure*}

\section{Discussion}

   We can now discuss the main results of our measurements on 2M-WS$_{2}$:

(i) The temperature dependence of the superfluid density, reflected by the one of the relaxation rate $\sigma_{\rm sc}$, is well described at all pressures by a single gap $s$-wave model. This is surprising since a multi-band bulk electronic structure with hole-like bands around the $N$ point and electron-like bands around the ${\Gamma}$ point was found for 2M-WS$_{2}$. On the other hand, Hall effect measurements showed that the hole-type carriers are dominant with a very high density of $n_{e,h}$ = 8.58 ${\times}$ 10$^{26}$ m$^{-3}$ at $T$ = 10 K. It is interesting that the superfluid density $n_{s}/(m^{*}/m_{e}$) ${\simeq}$ 2.8 ${\times}$ 10$^{26}$ m$^{-3}$, which we estimated for 2M-WS$_{2}$ is very close to the normal state hole density $n_{e,h}$, indicating that the superconductivity stems mostly from hole carriers. Thus, single gap superconductivity in this multi-band system may be explained by the fact that the superconducting gap occurs only on the hole-like Fermi surface and that the carrier density on the electron Fermi surface is not sufficient to induce superconductivity. This conclusion is also substantiated by the fact that superfluid density is insensitive to the shrinkage of the electron pocket, caused by pressure. If superconductivity would exist on the electron pocket, the pressure induced reduction of the size of the electron pocket would result to the reduction of the associated electron carrier concentration and thus the reduction of the superfluid density, which is not the case in 2M-WS$_{2}$. Single gap $s$-wave superconductivity in 2M-WS$_{2}$ is also different from the two-gap $s$+$s$-wave superconductivity, observed in the superconducting TMDs $T_{\rm d}$-MoTe$_{2}$ and 2H-NbSe$_{2}$. This again might be related to the fact that in both these systems both hole and electron bands contribute to superconductivity, while in 2M-WS$_{2}$ the superconductivity most likely arises predominantly from the hole like band. Regarding the surface states, a single Dirac cone at the ${\Gamma}$ point was reported. When the Fermi level is exactly at the Dirac point, only a single Fermi surface was obtained \cite{YFang}. A single Fermi surface at the surface and a dominant hole-like Fermi surface in the bulk is consistent with single gap superconductivity, observed in our study. It was also suggested that if the superconductivity is induced on this single Fermi surface, topological superconductivity is expected. As the muons penetrate deep into the material, we cannot obtain information on the superconductivity on the surface and about its topological nature. However, because of the semimetallic bulk electronic structure, the topological surface states are partially mixed with bulk states. Thus, the superconducting properties in the bulk could, to some extent, reflect the properties also on the surface.


(ii) We observed, for the first time, a strong negative pressure effect on the critical temperature $T_{c}$ and the superconducting gap ${\Delta}$. Both quantities decrease by ${\sim}$ 30 ${\%}$ under 1.9 GPa.  
Note that the superconducting transition temperature, estimated by the Allen and Dynes modified McMillan formula \cite{YFang}, was found to be ${\sim}$ 2 K, which is a factor of 4 smaller than the experimental value $T_{\rm c}$ ${\simeq}$ 8.5 K. This indicates that some dynamical and anharmonic effects on the electron-phonon coupling as well as interband scattering between electron and hole pockets are most likely the reasons for high-$T_{\rm c}$ in 2M-WS$_{2}$. According to our DFT calculations, the electron pocket around the ${\Gamma}$ point  shrinks by pressure, which would lead to the suppression of interband processes and this might have some effect on $T_{\rm c}$.  Pressure could also weaken the anharmonic electron-lattice interactions, giving rise to the suppression of  $T_{\rm c}$. Some additional experiments are certainly needed to understand the negative pressure effect on the critical temperature. The superfluid density $n_{s}/(m^{*}/m_{e}$) is essentially pressure independent, displaying at the most a slight increase of about 5${\%}$. The ratio between the superfluid density and the critical temperature  is located in the Uemura plot close to those of $T_{\rm d}$-MoTe$_{2}$ and 2H-NbSe$_{2}$ and other unconventional  superconductors, as shown in Figure 5. This suggests that 2M-WS$_{2}$ exhibits rather unconventional SC properties and might suggest a common mechanism and related electronic origin for the superconductivity in these superconducting TMDs. On the other hand, the insensitivity of the superfluid density to the strong change in $T_{c}$ is in contrast to the strong correlation between these two quantities in $T_{\rm d}$-MoTe$_{2}$ and 2H-NbSe$_{2}$. Let us try to understand this apparent discrepancy.
The nearly linear relationship between $T_{\rm c}$ and the superfluid density was originally observed in hole-doped cuprates in the underdoped region of the phase diagram \cite{Uemura1,Uemura3}, where the ratio between  $T_{\rm c}$ and their effective Fermi temperature $T_{\rm F}$ is about  $T_{\rm c}$/ $T_{\rm F}$ ${\sim}$ 0.05 corresponding to a 4-5 times reduction of $T_{\rm c}$ from the ideal Bose condensation temperature for a non-interacting Bose gas. These results were discussed in terms of a crossover  from Bose-Einstein condensation to BCS-like condensation \cite{Uemura4}. It is important to note that in cuprates the relationship between $T_{\rm c}$ and $n_{s}/(m^{*}/m_{e}$) was shown to depend on which part of the SC dome the samples are located. Thus, the left-side of the dome (underdoped), which has close proximity with the competing/cooperating antiferromagnetic phase, the center of the dome (optimally doped), which is away from the AFM phase or the right side of the dome (overdoped) region, which is clearly seperated from the AFM phase. The linear increase of  $T_{\rm c}$ with $n_{s}/(m^{*}/m_{e}$) is observed only in the underdoped region of the phase diagram. Within the optimal doping region $T_{\rm c}$ does not change despite the fact that $n_{s}/(m^{*}/m_{e}$) increases. In the overdoped region, $T_{\rm c}$ decreases, but $n_{s}/(m^{*}/m_{e}$) shows a small increase only. A similar behaviour was also observed in Fe-based superconductors \cite{Hashimoto}. With this in mind, we compare the temperature-pressure phase diagrams for the systems  2M-WS$_{2}$, $T_{\rm d}$-MoTe$_{2}$, 2H-NbSe$_{2}$.
The phase diagrams are shown in Fig. 5 (b-d). There are two obvious differences between the phase diagrams.  In the case of 2H-NbSe$_{2}$ and $T_{\rm d}$-MoTe$_{2}$ (Fig. 5 c,d), the phase diagram consists of two different orders at ambient pressure: superconductivity and either charge density wave order or non-centrosymmetric structural order. In addition, $T_{\rm c}$ of both systems increases with increasing pressure within the investigated pressure region (0 - 2 GPa), indicating that the samples are located in the left-hand side of the dome with close proximity to another phase. In the case of 2M-WS$_{2}$ (Fig. 5b), the system has solely superconducting order and there is no additional competing/cooperating order. In addition,  the system 2M-WS$_{2}$ has the highest $T_{\rm c}$ at ambient pressure, suggesting that the system is already in its optimal conditions in terms of superconductivity. Upon application of pressure, $T_{\rm c}$  decreases substantially, pushing the system towards the right-hand side of the dome, i.e. pushing it even further away from any possible competing phase.

In conclusion, using DFT, we show that the topology of the bands in 2M-WS$_{2}$  remains nontrivial  up to the highest applied pressure. We further provide the first microscopic investigation of the superconductivity in the layered superconductor 2M-WS$_{2}$ at ambient pressure as well as under hydrostatic pressures by means of muon-spin rotation. Specifically, the zero-temperature magnetic penetration depth ${\lambda}\left(0\right)$ and the temperature dependence of the superfluid density were studied in 2M-WS$_{2}$ by means of ${\mu}$SR experiments with pressures up to $p {\simeq}$ 1.9 GPa. The temperature dependence of the superfluid density ${\lambda^{-2}}$ is well described by a single gap $s$-wave scenario with a gap value of ${\Delta}$ ${\simeq}$ 1.31(1) meV at ambient pressure. Considering the previous observations of the dominant hole-like Fermi surface in the bulk and single Fermi surface on the surface we suggest that the well resolvable SC gap occurs only on the hole-like Fermi surface.
We also find that the ratio $T_{\rm c}$/$n_{\rm s}$ is located in the Uemura plot close to those of $T_{\rm d}$-MoTe$_{2}$ and 2H-NbSe$_{2}$ and other unconventional superconductors, suggesting that 2M-WS$_{2}$ exhibits non-standard SC properties. Moreover, we observed the strong negative pressure effect on $T_{\rm c}$ and ${\Delta}$, which may partly be due to the reduction of the size of the electron pocket at the ${\Gamma}$ point, at which a band inversion appears up to the highest applied pressure. The superfluid density shows only a very weak positive pressure effect and thus the correlation between the superfluid density and the critical temperature is absent in 2M-WS$_{2}$.  This is in contrast to the strong pressure effects on ${\lambda^{-2}}$ and its correlation with $T_{\rm c}$, observed in the superconducting TMDs $T_{\rm d}$-MoTe$_{2}$ and 2H-NbSe$_{2}$. This discrepancy is discussed in terms of the absence of competing or cooperating order to superconductivity in 2M-WS$_{2}$. These results hint towards the complex nature of the superconductivity in TMDs, despite its $s$-wave nature, which might have far reaching consequences for the future development of devices based on these materials.

\section{METHODS}

\textbf{Sample preparation}:  Topochemical synthesis procedure of the metastable, monoclinic tungsten diselenide (2M-WS$_2$) can be found elsewhere \cite{YFang,Wypych}. In a first step, a hexagonal phase of 2M-WS$_2$ (SG P6$_3$/mmc - No 194), was synthetized by solid state reaction method at 850$^{\circ}$C for 35 h from a stoichiometric mixture of elemental sulfur (5N, Alfa Aesar) and tungsten (5N, Alfa Aesar). Than potassium intercalation and single crystal growth of KWS$_2$ was conducted by annealing the stoichiometric mixture of K (3N5, Aldrich) and 2M-WS$_2$ at 850$^{\circ}$C. After 35 h a specimen was cooled down to 550$^{\circ}$C with a rate of 6$^{\circ}$C/h. All syntheses reactions and single crystal growth processes described above where conducted in a double wall, evacuated and sealed quartz ampules. In a third step K$_x$(H$_2$O)yWS$_2$ compound was created by washing ~3g of KWS$_2$ single crystals with deionized water. Finally K$_x$(H$_2$O)yWS$_2$ was oxidized with 200 ml of 0.1N K$_2$Cr$_2$O$_7$/0.1N H$_2$SO$_4$ aqueous solution in order to get 2M-WS$_2$.\\

\textbf{Pressure cell}: Double wall piston-cylinder type of cell made of CuBe/MP35N material was used in ${\mu}$SR experiments to generate pressures up to 1.9 GPa \cite{GuguchiaPressure,Andreica,MaisuradzePC,GuguchiaNature}. 
Small indium plate was placed together with the sample in the pressure cell filled with the Daphne oil. 
The pressure was estimated by tracking the SC transition of a indium plate by AC susceptibility.\\

\textbf{${\mu}$SR experiment}: Spin-polarized muons ${\mu}$$^{+}$ are extremely sensitive magnetic probes. 
Muon-spin experiences the Larmor precession either in the local field or in an applied magnetic field. By probing the field distribution in the vortex state of a Type-II superconductor using transverse field ${\mu}$SR technique, fundamental parameters such as the magnetic penetration depth $\lambda$ and the coherence length $\xi$ can be measured in the bulk of a superconductor. The analysis of TF-${\mu}$SR data \cite{Bastian,AndreasSuter} and the temperature dependence of the magnetic penetration depth \cite{carrington} is described in our previous work \cite{GuguchiaMoTe2}.\\

\textbf{DFT Calculations}: 2M-WS$_{2}$ crystal has a monoclinic structure with the space group symmetry $C_{2/m}$ (number 12) and the following lattice parameters $a = 12.856 \, \rm{\AA}$, $b = 3.2257 \, \rm{\AA}$, $c = 5.7109 \, \rm{\AA}$, $\alpha = \gamma = 90^{\circ}$ and $\beta = 112.942 \,^{\circ}$ (crystallographic data CCDC 1853656). 
 Full-relativistic electronic structure calculations were performed in Vienna \textit{ab initio} simulation package (VASP)~\cite{KresseFurth96-1, KresseFurth96-2} using Perdew-Burke-Ernzerhof (PBE)~\cite{PBE-1, PBE-2} generalized gradient approximation (GGA) exchange-correlation functional and projected augmented wave (PAW) potentials~\cite{PAW-1, PAW-2}. The van der Waals interactions were taken into account within the DFT-D2 approach by Grimme~\cite{Grimme}. The plane wave basis cutoff was set to 600 eV and $ 9 \times 9 \times 9$ k-points grids were generated within Monkhorst-Pack method.

\section{Acknowledgments}~
The ${\mu}$SR experiments were carried out at the Swiss Muon Source (S${\mu}$S) Paul Scherrer Insitute, Villigen, Switzerland using the HAL-9500 ${\mu}$SR spectrometer (${\pi}$E3 beamline), 
equipped with BlueFors vacuum-loaded cryogen-free dilution refrigerator (DR), GPS instrument (${\pi}$M3 beamline) and high pressure GPD instrument (${\mu}$E1 beamline). 
The work at the University of Zurich was supported by the Swiss National Science Foundation under Grant No. PZ00P2\_174015. R.K. acknowledges the Swiss National Science Foundation (grants 200021\_149486 and 200021\_175935). T.N. acknowledges the funding from the European Research Council (ERC) under the European Union`s Horizon 2020 research and innovation programm (ERC-StG-Neupert-757867-PARATOP). M.Z.H. is supported by US DOE/BES grant no. DE-FG-02-05ER46200. Z.G. acknowledges the valuable discussions with Jiaxin Yin.\\




\begin{thebibliography}{22}


\bibitem{Soluyanov} A. Soluyanov, et. al.,
Nature \textbf{527}, 495 (2015).

\bibitem{Kaminski} L. Huang, T.M. McCormick, M. Ochi, Z. Zhao, M.-T. Suzuki, R. Arita, 
Y. Wu, D. Mou, H. Cao, J. Yan, N. Trivedi, and A. Kaminski.
Nature Mater. \textbf{15}, 1155 (2016).

\bibitem{Xu} Xu, X., Yao, W., Xiao, D. and Heinz, T. F. 
Nat. Phys. \textbf{10}, 343 (2014).

\bibitem{Ali1} M.N. Ali, et. al., 
Nature 514, 205 (2014).

\bibitem{QiCava} Y. Qi $et~al.$, 
Nat. Comm. \textbf{7}, 11038 (2016). 

\bibitem{Qian} Qian, X., Liu, J., Fu, L. and Li, J. 
Science \textbf{346}, 1344-1347 (2014).

\bibitem{Zhang} Zhang, Y. J., Oka, T., Suzuki, R., Ye, J. T. and Iwasa, Y. 
Science \textbf{344}, 725-728 (2014).

\bibitem{Ugeda} M.M. Ugeda, A.J. Bradley, Y. Zhang, S. Onishi, Y. Chen, W. Ruan, C. Ojeda-Aristizabal, H. Ryu, M.T. Edmonds, H. Tsai, A. Riss, S.-K. Mo, D. Lee, A. Zettl, Z. Hussain, Z.-X. Shen, and M.F. Crommie, 
Nature Physics \textbf{12}, 92 (2016).

\bibitem{Bozin} Runze Yu, S. Banerjee, H.C. Lei, Ryan Sinclair, M. Abeykoon, H. D. Zhou, C. Petrovic, Z. Guguchia, and E. S. Bozin.
Phys. Rev. B \textbf{97}, 174515 (2018).

\bibitem{SunY} Sun, Y., Wu, S.C., Ali, M.N., Felser, C., and Yan, B. 
Phys. Rev. B \textbf{92}, 161107 (2015).

\bibitem{Hasan1} I. Belopolski, D.S. Sanchez, Y. Ishida, X. Pan, P. Yu, S.-Y. Xu, G. Chang, T.-R. Chang, H. Zheng, N. Alidoust, G. Bian, M. Neupane, S.-M. Huang, C.-C. Lee, Y. Song, H. Bu, G. Wang, S. Li, G. Eda, H.-T. Jeng, T. Kondo, H. Lin, Z. Liu, F. Song, S. Shin, and M.Z. Hasan.
Nature Communications \textbf{7}, 13643 (2016).

\bibitem{Le} L.P. Le, G.M. Luke, B.J. Sternlieb, W.D. Wu, Y.J. Uemura, J.W. Brill, and H. Drulis,
Physica C \textbf{185}, 2715 (1991).

\bibitem{KivelsonCDW} Yue Yu and S. A. Kivelson,
https://arxiv.org/abs/1809.10160.

\bibitem{Yokoya} T. Yokoya, et. al., 
Science \textbf{294}, 2518 (2001).

\bibitem{Sipos} B. Sipos, A. F. Kusmartseva, A. Akrap, H. Berger, L. Forro, and E. Tutis, Nat. Mat. \textbf{7}, 960 (2008).

\bibitem{Tidman} J.P. Tidman, O. Singh, A.E. Curzon, R.F. Frindt, Phil. Mag. \textbf{30}, 1191 (1974).

\bibitem{Thompson} A.H. Thompson, F.R. Gamble, R.H. Koehler Jr., Phys. Rev. B \textbf{5} 2811 (1972).

\bibitem{Naito} M. Naito, S. Tanaka, J. Phys. Soc. Jpn. \textbf{51}, 219 (1982).

\bibitem{Iwasa2012} J. T. Ye, Y. J. Zhang, R. Akashi, M.S. Bahramy, R. Arita, and Y. Iwasa,
Science \textbf{338}, 1193 (2012).

\bibitem{Iwasa2015} W. Shi, J. Ye, Y. Zhang, R. Suzuki, M. Yoshida, J. Miyazaki, N. Inoue, Y. Saito and Y. Iwasa, 
Scientific Reports \textbf{5}, 12534 (2015).

\bibitem{Klemm}  R.A. Klemm.
Physica C \textbf{514}, 86-94 (2015).

\bibitem{JLu}  J. Lu, O. Zheliuk, Q. Chena, I. Leermakers, N.E. Hussey, U. Zeitler, and J. Ye.
PNAS \textbf{115}, 3551-3556 (2018).


\bibitem{GuguchiaMoTe2} Z. Guguchia, F. von Rohr, Z. Shermadini, A. T. Lee, S. Banerjee, A. R. Wieteska, C. A. Marianetti, B. A. Frandsen, H. Luetkens, Z. Gong, S. C. Cheung, C. Baines, A. Shengelaya, G. Taniashvili, A. N. Pasupathy, E. Morenzoni, S. J. L. Billinge, A. Amato, R. J. Cava, R. Khasanov, Y. J. Uemura, 
Nature Communications \textbf{8}, 1082 (2017).

\bibitem{GuguchiaNbSe2}  F. von Rohr, J.-C.~Orain, R.~Khasanov, Z. Shermadini, A.~Nikitin, J.~Chang, A.R.~Wieteska, A.N.~Pasupathy, M.Z.~Hasan, A.~Amato, H.~Luetkens, Y. J. Uemura, and Z. Guguchia. 
arXiv:1903.05292 (2019).

\bibitem{Uemura1} Y.J. Uemura et. al., Phys. Rev. Lett. \textbf{62}, 2317 (1989).

\bibitem{Uemura3} Y.J. Uemura et. al., Nature \textbf{352}, 605 (1991).

\bibitem{Uemura7}  Y.J. Uemura, 
Nature Materials \textbf{8}, 253-255 (2009).

\bibitem{Uemura4} Y.J. Uemura et. al., 
Phys. Rev. Lett. \textbf{66}, 2665 (1991).

\bibitem{EmeryKivelson} V. Emery and S. Kivelson, 
Nature \textbf{374}, 434 (1995).

\bibitem{Shengelaya} A. Shengelaya, R. Khasanov, D.G. Eshchenko, D. Di Castro,
I.M. Savic, M.S. Park, K.H. Him, Sung-IK Lee, K.A. M\"{u}ller, and H. Keller, Phys. Rev. Lett. \textbf{94}, 127001 (2005).

\bibitem{Luetkens} H. Luetkens, H.-H. Klauss, M. Kraken, F.J. Litterst, T.
Dellmann, R. Klingeler, C. Hess, R. Khasanov, A. Amato,
C. Baines, M. Kosmala, O.J. Schumann, M. Braden,
J. Hamann-Borrero, N. Leps, A. Kondrat, G. Behr, J.
Werner, and B. B\"{u}chner, 
Nature Materials \textbf{8}, 305 (2009).

\bibitem{Carlo} J.P. Carlo, Y.J. Uemura, T. Goko, G.J. MacDougall, J.A. Rodriguez, W. Yu, G.M. Luke, P. Dai, N. Shannon, S. Miyasaka, S. Suzuki, S. Tajima, G.F. Chen, W.Z. Hu, J.L. Luo, and N.L. Wang, Phys. Rev. Lett. \textbf{102}, 087001 (2009).

\bibitem{Khasanov2008} R. Khasanov, H. Luetkens, A. Amato, H.-H. Klauss, Z.-A. Ren, 
J. Yang, W. Lu, and Z.-X. Zhao, Phys. Rev. B \textbf{78}, 092506 (2008).

\bibitem{GuguchiaPRB} Guguchia, Z., Shermadini, Z., Amato, A., Maisuradze, A.,
Shengelaya, A., Bukowski, Z., Luetkens, H., Khasanov, R., Karpinski, J.
and Keller, H. 
$Phys.~Rev.~B$ \textbf{84}, 094513 (2011).

\bibitem{YFang}  Y. Fang, J. Pan, D. Zhang, D. Wang, H.T. Hirose, T. Terashima, S. Uji, Y. Yuan, W. Li, Z. Tian, J. Xue, Y. Ma, W. Zhao, Q. Xue, G. Mu, H. Zhang, and F. Huang.
arXiv:1808.05324 (2018).

\bibitem{Hosur} Hosur, P., Dai,  X., Fang, Z., and Qi, X.-L.,
Phys. Rev. B \textbf{90}, 045130 (2014).

\bibitem{Ando}  Ando, Y. and Fu, L.
Annual Review of Condensed Matter Physics \textbf{6}, 361 (2015).

\bibitem{Grushin} Grushin, A.G. Phys. Rev. D \textbf{86}, 045001 (2012).

\bibitem{GuguchiaPressure} R. Khasanov, Z. Guguchia, A. Maisuradze, D. Andreica, M. Elender, A. Raselli, Z. Shermadini, T. Goko, F. Knecht, E. Morenzoni, and A. Amato,  
High Pressure Research \textbf{36}, 140-166 (2016).

\bibitem{Andreica} Andreica D 2001 $Ph.D.~thesis$ IPP/ETH-Z\"{u}rich.

\bibitem{MaisuradzePC} A. Maisuradze, A. Shengelaya, A. Amato, E. Pomjakushina, and H. Keller
Phys. Rev. B \textbf{84}, 184523 (2011).

\bibitem{GuguchiaNature} Z. Guguchia, A. Amato, J. Kang, H. Luetkens, P.K. Biswas, G. Prando, F. von Rohr, Z. Bukowski, A. Shengelaya, H. Keller, E. Morenzoni, R.M. Fernandes, and R. Khasanov, 
Nature Communications \textbf{6}, 8863 (2015).

\bibitem{Sonier} J. E. Sonier, J.H. Brewer, and R.F. Kiefl, 
Rev. Mod. Phys. \textbf{72}, 769 (2000).

\bibitem{Toyabe} R. Kubo and T. Toyabe, 
(North Holland, Amsterdam, 1967).

\bibitem{GuguchiaPRB2016}  Z. Guguchia, R. Khasanov, Z. Bukowski, F. von Rohr, M. Medarde, P.K. Biswas, H. Luetkens, A. Amato, and E. Morenzoni,
Phys. Rev. B \textbf{93}, 094513 (2016).

\bibitem{KhasanovPRL103} R. Khasanov, A. Maisuradze, H. Maeter, A. Kwadrin, H. Luetkens, A. Amato, W. Schnelle, H. Rosner, A. LeitheJasper, and H.-H. Klauss, 
Phys. Rev. Lett. \textbf{103}, 067010 (2009).

\bibitem{Williams2010} T.J. Williams, A.A. Aczel, E. Baggio-Saitovitch, S.L. Bud’ko, P.C. Canfield, J.P. Carlo, T. Goko, H. Kageyama, A. Kitada, J. Munevar, N. Ni, S.R. Saha, K. Kirschen-baum, J. Paglione, D.R. Sanchez-Candela, Y.J. Uemura, and G. M. Luke, 
Phys. Rev. B \textbf{82}, 094512 (2010).

\bibitem{SonierPRL2011} J.E. Sonier, W. Huang, C.V. Kaiser, C. Cochrane, V. Pacradouni, S.A. Sabok-Sayr, M.D. Lumsden, B.C. Sales, M.A. McGuire, A.S. Sefat, and D. Mandrus, 
Phys. Rev. Lett. \textbf{106}, 127002 (2011).

\bibitem{Brandt} Brandt, E.H. 
$Phys.~Rev.~B$ \textbf{37}, 2349 (1988).

\bibitem{Tinkham} Tinkham, M. 
$Krieger~Publishing~Company$, $Malabar,~Florida$, 1975.

\bibitem{Frandsen} B.A. Frandsen, S.C. Cheung, T. Goko, L. Liu, T. Medina, T.S.J. Munsie,
G.M. Luke, P.J. Baker, M.P. Jimenez S., G. Eguchi, S. Yonezawa, Y. Maeno, and Y.J. Uemura,
Phys. Rev. B \textbf{91}, 014511 (2015).

\bibitem{Hashimoto} K. Hashimoto, K. Cho, T. Shibauchi, S. Kasahara, Y. Mizukami, R. Katsumata, Y. Tsuruhara, 
T. Terashima, H. Ikeda, M.A. Tanatar, H. Kitano, N. Salovich, R.W. Giannetta, P. Walmsley, A. Carrington, 
R. Prozorov, and Y. Matsuda.
Science \textbf{336}, 1554-1557 (2012).

\bibitem{Wypych} F. Wypych, K. Sollmann, and R. Schollhorn,
Materials Research Bulletin \textbf{27}, 545-553 (1992).

\bibitem{Bastian} Suter, A. and Wojek, B.M. $Physics~Procedia$ \textbf{30}, 69 (2012).\\
The fitting of the $T$-dependence of the penetration depth with
${\alpha}$ model was performed using the additional library BMW developped
by B.M. Wojek.

\bibitem{AndreasSuter} Suter, A. and Wojek, B.M. 
$Physics~Procedia$ \textbf{30}, 69-73 (2012).

\bibitem{carrington} Carrington, A. and Manzano, F. 
$Physica~C$ \textbf{385}, 205 (2003).

\bibitem{KresseFurth96-1}  G. Kresse and J. Furthm\"uller,
Computational Materials Science \textbf{6}, 15 (1996).

\bibitem{KresseFurth96-2} G. Kresse and J. Furthm\"uller,
Phys. Rev. B \textbf{54}, 11169 (1996).

\bibitem{PBE-1} J.P. Perdew, K. Burke, and M. Ernzerhof, 
Phys. Rev. Lett. \textbf{77}, 3865 (1996).

\bibitem{PBE-2} J. P. Perdew, K. Burke, and M. Ernzerhof, 
Phys. Rev. Lett. \textbf{78}, 1396 (1997).

\bibitem{PAW-1} P.E. Bl\"ochl, 
Phys. Rev. B \textbf{50}, 17953 (1994).

\bibitem{PAW-2} G. Kresse, and D. Joubert,
Phys. Rev. B \textbf{59}, 1758 (1999).

\bibitem{Grimme} S. Grimme,
Journal of Computational Chemistry \textbf{27}, 1787 (2006).

\bibitem{YFeng} Y. Feng, J. Wang, R. Jaramillo, J. van Wezel, S. Haravifard, G. Srajer, Y. Liu, Z.-A. Xu,
P.B. Littlewood, and T. F. Rosenbaum,
Proc. Natl. Acad. Sci. \textbf{109}, 7224-7229 (2012).




\end{thebibliography}
\end{document}